\documentclass{aa} %

\usepackage{graphicx}
\usepackage{txfonts}
\usepackage[table]{xcolor} 

\newcommand{\markme}[1]{\textbf{#1}}
\usepackage{float}
\usepackage{tabularx}
\usepackage{ulem}
\usepackage{gensymb}

\newcommand{\B}{B}
\newcommand{\Tp}{T_{\textrm{p}}}
\newcommand{\Sp}{S_{\textrm{p}}}
\newcommand{\n}{n_{\textrm{p}}}

\newcommand{\vp}{v_{\textrm{p}}}
\newcommand{\va}{v_{\textrm{A}}}
\newcommand{\ac}{a_{\textrm{col,p-p}}}
\newcommand{\nO}{n_{O^{7+}}/n_{O^{6+}}}
\newcommand{\nC}{n_{C^{6+}}/n_{C^{5+}}}
\newcommand{\Tr}{T_{\text{rel}}}

\begin{document} 

\title{Proton-proton collisional age to order solar wind types}
\titlerunning{}
\author{Verena Heidrich-Meisner \inst{1}, Lars Berger \inst{1}, and Robert F. Wimmer-Schweingruber \inst{1}}

\authorrunning{V. H.-M., T. P., M. K., L. B.,  R. W.-S.}

\institute{Christian Albrechts University at Kiel, Germany,
  \email{heidrich@physik.uni-kiel.de}
}

\date{}

\abstract{
The properties of a solar wind stream are determined by its source region and by transport effects. Independently of the solar wind type, the solar wind measured in situ is always affected by both. This means that reliably determining the solar wind type from {in situ} observations is useful for the analysis of {its solar origin} and its evolution during the travel time to the spacecraft that observes the solar wind. In addition, the solar wind type also influences the interaction of the solar wind with other plasma such as Earth's magnetosphere. 
}
{
We consider the proton-proton collisional age as an ordering parameter for the solar wind at 1AU and explore its relation to the solar wind classification scheme developed by \citet[]{xu2014new}. {We use this to show that explicit magnetic field information is not required for this solar wind classification.} Furthermore, we illustrate that solar wind classification schemes that rely on {threshold values of} solar wind parameters should depend on the phase in the solar activity cycle since the respective parameters change with the solar activity cycle.
}
{
  The categorization of the solar wind follwing \citet{xu2014new} was taken as our reference for determining the solar wind type{. B}ased on the observation that the three basic solar wind types from this categorization cover different regimes in terms of proton-proton collisional age $\ac$, we propose a simplified solar wind classification scheme that is only based on the proton-proton collisional age. {We call the resulting method the PAC solar wind classifier.} {For this purpose}, we derive time-dependent threshold values {in} the proton-proton collisional age {for two variants of the proposed PAC scheme: (1) similarity-PAC is based on the similarity to the full \citet[]{xu2014new} scheme, and (2) distribution-PAC is based directly on the distribution of the proton-proton collisional age.}  
}
{
{The proposed simplified solar wind categorization scheme based on the proton-proton collisional age represents an equivalent alternative to the full \citet{xu2014new} solar wind classification scheme and leads to a classification that is very similar to the full \citet[]{xu2014new} scheme. The proposed PAC solar wind categorization} separates coronal hole wind from helmet-streamer plasma as well as helmet-streamer plasma (slow solar wind without a current sheet crossing) from sector-reversal plasma (slow solar wind with a current sheet crossing). {Unlike the full \citet{xu2014new} scheme, PAC does not require information on the magnetic field as input.}
}
{
The solar wind is well ordered by the proton-proton collisional age. This implies underlying intrinsic relationships between the plasma properties, in particular, proton temperature and magnetic field strength in each plasma regime.  {We argue that sector-reversal plasma is a combination of particularly slow and dense solar wind and most stream interaction boundaries.} Most solar wind parameters (e.g., {the} magnetic field strength{,} $B$, and {the oxygen charge state ratio} $\nO$) change with the solar activity cycle. Thus, all solar wind categorization schemes based on threshold values need to be adapted to the solar activity cycle as well. {Because it does not require magnetic field information but only proton plasma measurements, the proposed PAC solar wind classifier can be applied directly to solar wind data from the Solar and Heliospheric Observatoty (SOHO), which is not equipped with a magnetometer.}
}


   \maketitle
%

   \section{Introduction}
   Historically, the variable plasma of the solar wind has been
   categorized into two distinctly different regimes: coronal hole
   wind, and slow solar wind
   \citep[]{mccomas2000solar,vonSteiger2000,zhao2010comparison,xu2014new}.
   The simplistic view of mainly two types of solar wind has   frequently been challenged
   \citep[]{stakhiv2015origin,d2015origin,sanchez2016very}. {Results}
   of an unsupervised machine-learning approach based on both charge
   state composition and proton plasma properties indicate
   seven types of solar wind \citep[]{heidrich2018solar}.
   
   The {in situ} measured properties of the solar wind are determined by
   the respective solar source region and by transport processes. Most
   solar wind categorization schemes implicitly assume that transport
   effects are negligible. Different solar wind types (e.g.,
   coronal hole wind and slow solar wind) are associated with
   different solar origins and release mechanisms. Plasma from stream
   interaction regions {is in addition} strongly affected by
   transport processes.

   Solar wind categorization schemes rely on different solar wind
   properties to {identify} the solar wind type and adopt
   mainly one of the following two approaches: 1) Composition-based
   schemes exploit the {oxygen} and {carbon}
   charge-state composition of the solar wind
   \citep[]{zhao2010comparison,vonSteiger2000}. {Based on
     the assumption that the charge states observed in the solar wind
     are determined in the solar corona and are not significantly
     changed thereafter, l}ower (higher) charge states {are
     associated with source region{s} of the respective
     solar wind stream with comparatively low (high) electron
     temperatures.} Because the charge state is not expected to change
   during the travel time of the solar wind, composition-based
   criteria are well suited to identify different solar source
   regions.  Transport effects{ due to, for instance,}
   compression regions in stream interaction regions and wave-particle
   {interaction} are not directly reflected in the charge-state composition. Stream interaction regions tend to be
   characterized by a (gradual or abrupt) transition in the
   {oxygen} charge-state compositions. {From the
     charge-state information alone, stream interaction regions cannot
     be uniquely identified, and their solar source regions cannot be
     unambiguously determined without additional or context
     information.}  2) Proton plasma properties provide an alternative
   to determine the solar wind type
   \citep[]{xu2014new,camporeale2017classification}. A clear advantage
   of this approach is that the required observables are available
   from more spacecraft. Unlike the charge-state composition, the
   proton speed, proton density, proton temperature, and magnetic
   field strength (and derived quantities such as the specific entropy
   and the Alfv{\'e}n speed) are all susceptible to transport
   effects. In particular, these quantities show radial gradients
   throughout the heliosphere
   \citep[]{marsch1982solar,bale2019highly,kasper2019alfvenic}. Thus,
   a solar wind categorization based on threshold values for these
   quantities can be expected to depend on position.  {In
     particular, the solar wind proton temperature is not a tracer of
     the coronal (electron) temperature. The
     solar wind proton temperatures show the opposite effect
     \citep[]{vonSteiger2000}: high solar wind proton temperatures are
     observed for coronal hole wind (which originates from
     comparatively cool coronal regions), while low solar wind proton
     temperatures appear in the slow solar wind (which likely originates
     in hot coronal regions). The solar wind proton temperature is
     probably strongly influenced by transport effects, in
     particular, by wave-particle interactions.}  In addition, proton
   temperature, proton density, and magnetic field strength all show
   characteristic variations in stream interaction regions. Solar wind categorization schemes based on proton plasma properties
   are therefore well suited to assess the effect of solar wind evolution during
   its travel time. However, these transport effects can blur the
   tracers of the solar origin of the solar wind. For approaches based on charge
-state composition and on proton plasma, the
   respective threshold values are usually determined heuristically
   and vary in the literature.  {Mainly as a result of their
     availability, solar wind electron data (e.g.,
     \citet[]{lin1995three,wilson2018statistical}) are typically not
     considered for solar wind classifications, although their
     properties, for example, the electron temperature and the
     electron-proton collisional age, can be expected to be
     informative in this context. Future improvements on solar wind
     classification would most likely benefit considerably from
     including electron data.}
    The collisional age (or {Coulomb} number) has been
    proposed as an ordering parameter for the solar wind in
    \citet{kasper2008hot},\citet{tracy2016constraining}, and \citet{maruca2013collisional}. The
    collisional age can be interpreted as counting the number of
    90$^{\degree}$-equivalent collisions during the travel time from
    the Sun to the observing spacecraft. {This notion of the
      collisional age relies on the simplifying assumption that the
      solar wind parameters are constant during the solar wind travel
      time.}  {\citet[]{maruca2013collisional} introduced an
      improvement in the computation of the collisional age that
      takes into account that the underlying quantities are not
      constant during the travel time of the solar wind.}

   We here investigate the relationship between the method developed by
   \citet{xu2014new} and the proton-proton collisional
   age. {We illustrate that simple thresholds in the
     proton-proton collisional age are sufficient to derive a new
     solar wind categorization that leads to a
     categorization that is very similar to the full \citet[]{xu2014new} scheme but does not require the magnetic field strength as input. Our
   proposed solar wind categorization scheme is called PAC (for
   proton-proton collisional age $\ac$) solar wind categorization. We derive
   two variants of the PAC scheme: one that is based on the similarity
   to the \citet[]{xu2014new} scheme, and a second that is based
   directly on the observed distribution of the proton-proton
   collisional age.}

   Our test solar wind data set is specified in
   {Sec.}~\ref{sec:ace}. We briefly describe the \citet[]{xu2014new}
   solar wind classification scheme and discuss examples of occasional
   {misclassifications} in {Sec.} \ref{sec:xu}. In {Sec.}
   \ref{sec:ac} we then show that the \citet[]{xu2014new} types are
   ordered by the proton-proton collisional age{,} investigate
   the underlying reason{s} for this,{ and derive the first version
     of our proposed solar wind categorization based on the similarity
     to the \citet[]{xu2014new} scheme.} Furthermore, in
   {Sec.}~\ref{sec:solarcycle} the evolution of typical solar wind
   parameters throughout the solar activity cycle is illustrated, and
   in {Sec.}~\ref{sec:time} we derive \markme{a second
   time-dependent version of our proposed solar wind classification method that is based on the distribution of the proton-proton collisional age.}

\begin{figure*}
  \includegraphics[width=\textwidth]{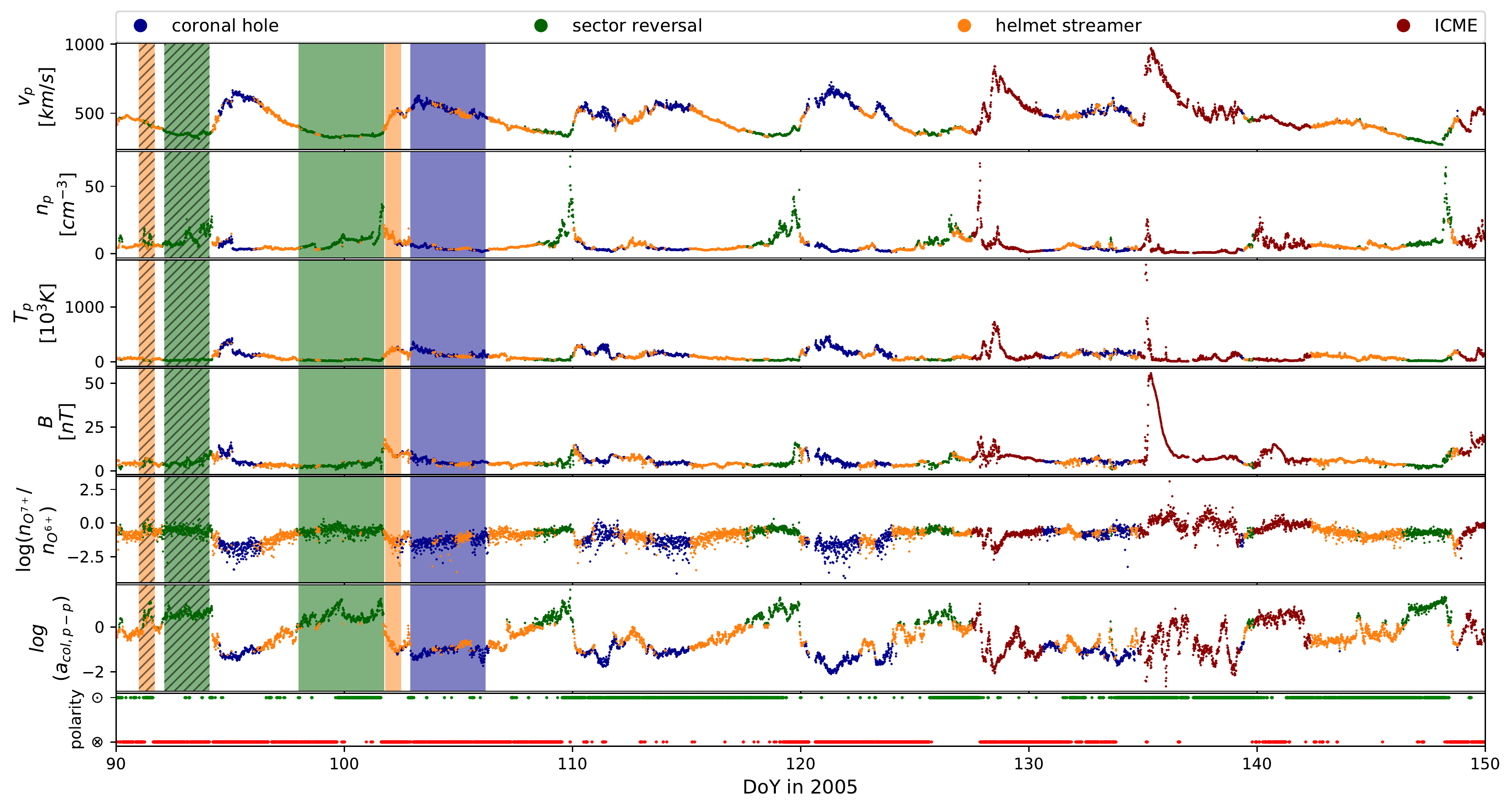}
  
  \caption{\label{fig:timeseries} Time series of solar wind parameters
    for day of year (DoY) 90-150 in 2005. The following solar wind
    parameters are shown (from top to bottom panel): proton speed
    $\vp$, proton density $\n$, proton temperature $\Tp$, magnetic
    field strength $\B$, {oxygen} charge-state ratio $\nO$, (decadic) logarithm of the
    proton-proton collisional age $\log{\ac}$, and in situ magnetic
    field polarity. The colored shading highlights different periods of
    interest. The color of the data points indicate{s} the solar wind
    type based on the \citet[]{xu2014new} scheme. ICMEs are based on
    available ICME lists. The in situ magnetic field polarity is
    determined by comparing the nominal magnetic field direction based on the
    Parker spiral to the in situ measured angle. If the absolute
    difference between these two directions is greater than $90$ degrees, the magnetic polarity is inwardly (crossed circle), red)
    directed, otherwise{,} it is outwardly (dotted circle, green) directed.}
\end{figure*}

\begin{figure*}\begin{center}
  \includegraphics[width=0.95\textwidth]{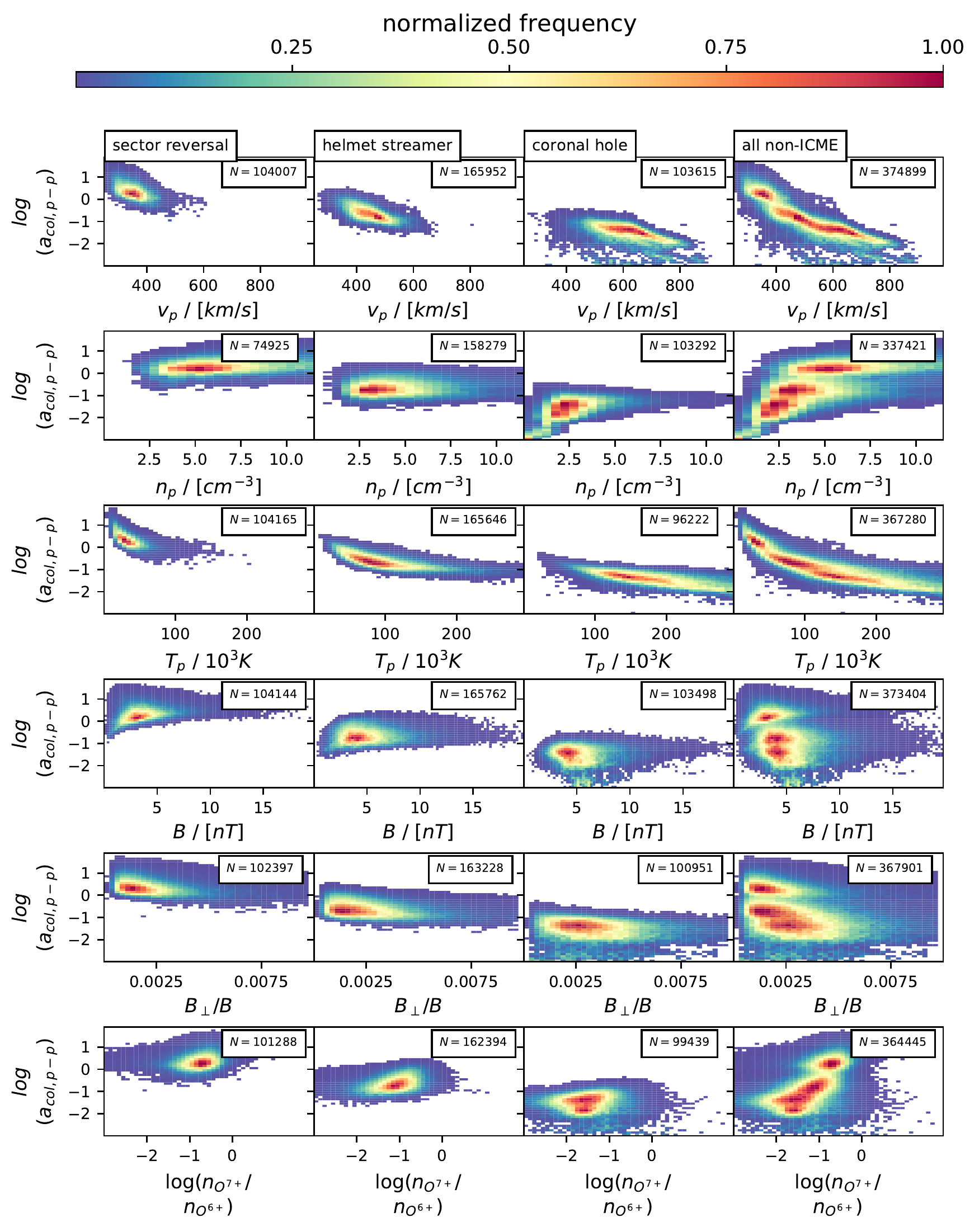}
  \end{center}
\caption{\label{fig:cols} \markme{two-dimensional} histograms of logarithmic proton-proton
    collisional age vs. solar wind parameters. Each histogram is
    normalized to its maximum value, {and each histogram values is
      divided by the respective bin size before normalization}. Each
    row corresponds to the same solar wind parameter (from top to
    bottom: proton speed $\vp$, proton temperature $Tp$, proton
    density $n$, magnetic field strength $\B$, perpendicular
    variability of the magnetic field $B_\perp/B$, {and oxygen}
    charge-state ratio $\nO$ ). The first three columns refer to the
    three different \citet{xu2014new} solar wind types (from left to
    right: sector-reversal plasma, helmet-streamer plasma, and coronal
    hole wind). The fourth column is the sum of the three other
    normalized histograms. {In each panel, the sample size ($N$)
      is given as an inset.}}
\end{figure*}

\begin{figure}
  \includegraphics[width=\columnwidth]{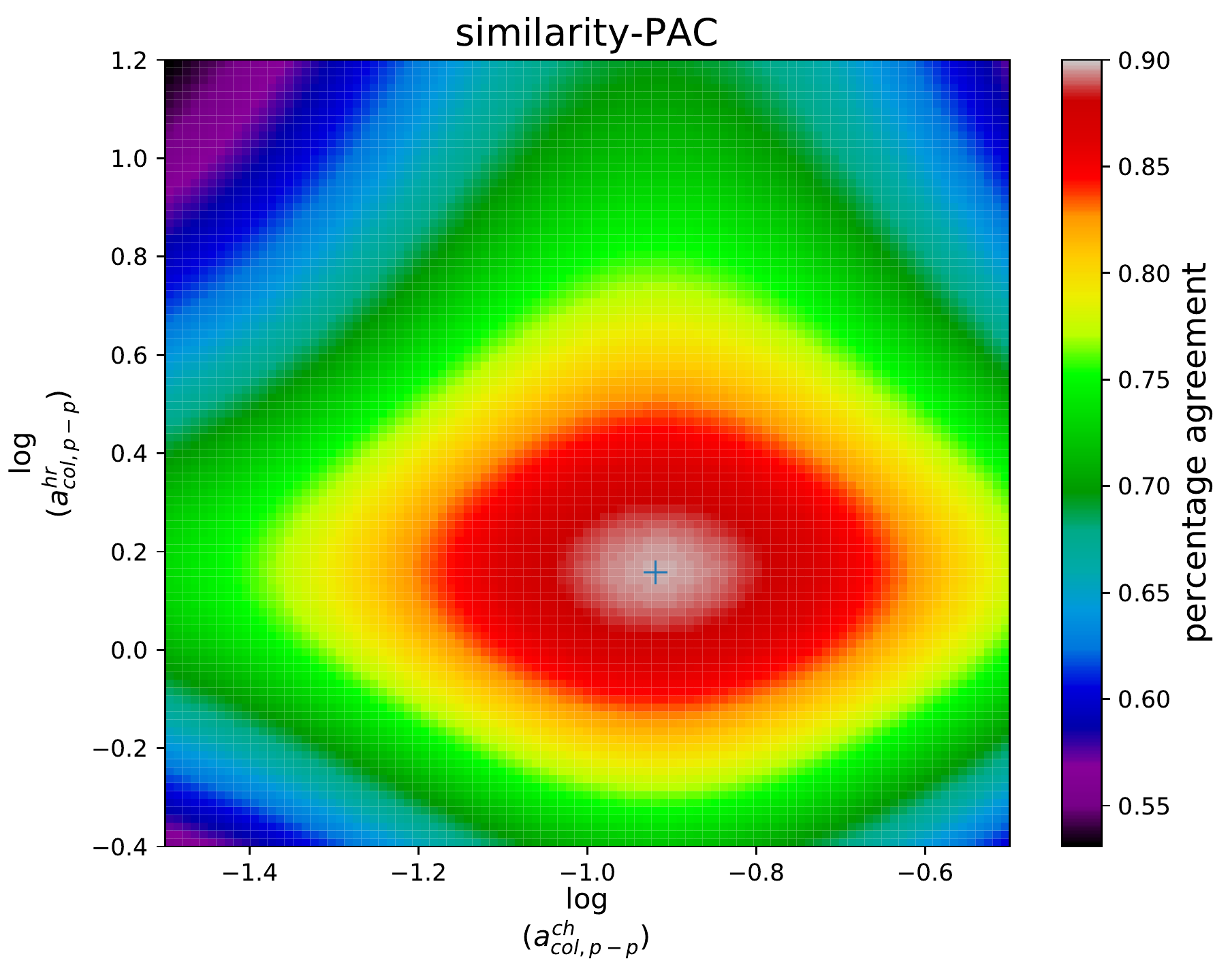}
  \caption{\label{fig:similarity} Similarity between {the} \citet{xu2014new} solar wind types and {the similarity-PAC} solar wind categorization for different threshold values. The x-axis varies the value of the threshold on the proton-proton collisional age between coronal hole wind and helmet-streamer plasma  {$\ac^{ch}$}, and the y-axis varies threshold values between helmet-streamer and sector-reversal plasma  {$\ac^{hr}$}. The similarity of the two categorizations is estimated as the ratio of the number of data points that are categorized as the same solar wind type for both schemes over  all valid (non-ICME) data points. The $\text{plus}$  indicates the maximum of the similarity measure.} 

\end{figure}

\begin{figure*}\begin{center}
  \includegraphics[width=0.95\textwidth]{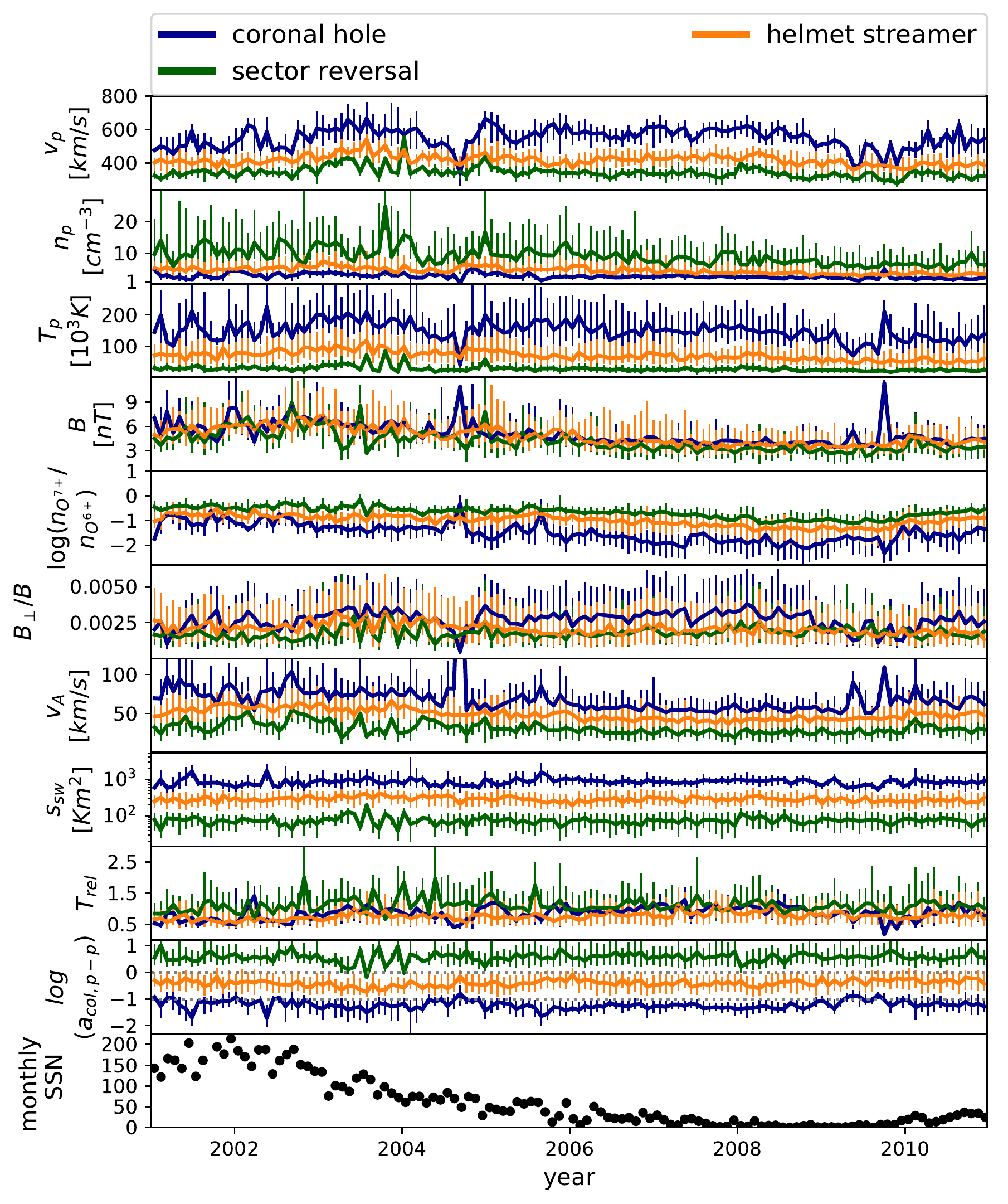}
  \end{center}
  \caption{\label{fig:CRTS} Time series of solar wind parameters averaged over {$27.24$ days} and separated by solar wind type in the \citet[]{xu2014new} scheme within each {time bin}. From top to bottom: Proton speed $\vp$,proton density $n$, proton temperature $\Tp$, magnetic field strength $\B$, {oxygen} charge-state ratio $\nO$, perpendicular variability of the magnetic field $\Delta B_\perp/B$,  Alfv{\'e}n speed $\va$, specific proton entropy $\Sp$, temperature ratio $\Tr$, proton-proton collisional age $\ac$ , and monthly SSN. Each point represents the median of all observations within {the respective time bin} with the selected solar wind type. The error bars indicate a confidence interval from the 15.9th- to the 84.1st percentile {(analogous to a $1\sigma$ confidence interval)}. As in {Fig.}~\ref{fig:timeseries}, the color indicates the solar wind type.}
\end{figure*}

\section{Test data: Ten years of observations from the Advanced Composition Explorer}\label{sec:ace}
As in \citet[]{xu2014new}, our considerations are based on observations
{from} the Advanced Composition Explorer (ACE, which orbits the first Lagrange point) as a test case for categorizing the solar wind. The instruments on ACE provide both proton plasma
properties and charge-state compositions. The magnetic field strength
($\B$) is taken from ACE/MAG \citep[]{smith1998ace}, the compositional
information (the ratios $\nO$ and $\nC$ based on the ion
densities $n_{O^{6+}}, n_{O^{7+}}, n_{C^{5+}}, n_{C^{6+}}$) from the
Solar Wind Ion Composition Spectrometer (SWICS,
\citet[]{gloeckler-etal-1998}), and the proton plasma properties
(proton speed $\vp$, proton density $\n$, and proton temperature
$\Tp${)} from a combined data set of the Solar Wind Electron
Proton and Alpha Monitor (SWEPAM, \citet[]{mccomas1998solar}) and
ACE/SWICS. Because the operational state of ACE/SWICS was altered due to
radiation and an age-induced hardware anomaly in August 2011 and
ACE/SWEPAM suffers from an increasing number of data gaps in later
years, we restrict our analysis to 2001-2010. To reduce the number of
data gaps, we used the combined SWEPAM-SWICS data set for the proton
plasma properties, although this restricts us to the 12-minute time
resolution of ACE/SWICS. The ACE/SWICS {heavy-}ion densities
are based directly on the ACE/SWICS pulse-height analysis data
\citep{berger2008velocity}. Because we are interested in the background
solar wind, we excluded interplanetary coronal mass ejections
({ICME}s) from our data set based on the Jian and
Richardson\&Cane ICME lists
\citep[]{jian2006properties,jian2011comparing,cane2003interplanetary,richardson2010near}. {A
  six-hour grace period before and after every ICME from either list
  was taken into account for possible imprecise start or end times.} The ejecta
category of the \citet[]{xu2014new} scheme was ignored. Instead, the
respective data points that are outside the ICMEs identified by the lists described above were added to the corresponding solar wind types
(i.e., we ignored the decision boundary Q1 from
\citet[]{xu2014new}). Following \citet{kasper2008hot}, the
proton-proton collisional age $\ac$ was derived from the proton plasma
properties
\begin{eqnarray}
  \ac = \frac{6.4 10^8K^{3/2}}{cm^{-3}}   \frac{\n}{\vp\Tp^{3/2}} \enspace , \label{eq:col}  
\end{eqnarray}
{with $v_p$ in km/s, $n_p$ in cm$^{-3}$, and $T_p$ in K.}
To obtain a measure of the wave activity, we derived the mean direction
over 12 minutes of the magnetic field vector, rotated the coordinate
system in this direction, and computed from the 1s time resolution data
from {ACE/}MAG the length of the two perpendicular components
of $B$ as $B_\perp=\sqrt{B_{\perp,x}^2+B_{\perp,y}^2}$ and average
$B_\perp$ over 12 minutes to determine $B_\perp/B$.

\section{\cite{xu2014new} solar wind categorization scheme}\label{sec:xu}

The solar wind {categorization} scheme proposed in \citet{xu2014new}
considers four solar wind types: ejecta (ICME plasma), coronal hole
wind, helmet-streamer plasma (slow solar wind without a current sheet
crossing), and sector-reversal plasma (slow solar wind around a current
sheet crossing). Here, we are interested in the background solar wind
and therefore disregard the ejecta category. Instead, {as
  described in the previous section,} ICMEs are excluded from the data set based on the
combination of the \citet{jian2006properties,jian2011comparing} and
\citet{richardson2010near} ICME lists. The \citet{xu2014new} scheme
places three separating planes in the \markme{three-dimensional} space spanned
by Alfv{\'e}n speed $\va$, specific proton entropy $\Sp$, and the
ratio $\Tr=\Tp/T_{\text{exp}}$ between the observed proton temperature
$\Tp$ and an expected proton temperature {(in
  eV, \footnote{This is the only time in this manuscript that eV and not K is
    used as unit for the temperature.})}
$T_{\text{exp}}=\vp/258^{3.113}$ {depending on the solar wind  proton speed $\vp$ in km/s}. An improved version
(\citet{camporeale2017classification} of the categorization scheme
trains a Gaussian process based on the same hand-selected plasma data
as in \cite{xu2014new}. For convenience, \citet{xu2014new} also
{provided} an expression for the decision boundaries based on proton
speed, proton density, proton temperature, and magnetic field
strength. Because we do not use the ejecta category itself, we did
not apply the decision boundary either (Q1 in \citet[]{xu2014new}), which only
serves the purpose of separating ejecta from the three other solar
wind types.

Figure \ref{fig:timeseries} shows an example of a time series of the
proton solar wind plasma properties. In each panel, the color
indicates the solar wind types assigned by the \citet{xu2014new}
scheme. For the most part, the resulting solar wind categorization
follows expectations very well. {Solar wind streams with
  low proton density, high proton temperatures, and high proton speeds
    are recognized as coronal hole wind, while low proton speeds, high
    densities, and low proton temperatures are associated with the two
    slow solar wind categories. }

  Stream interaction regions are sometimes assigned to either of the
  two slow solar wind types (because most but not all of them are
  associated with current {sheet} crossings). For example, the
  extended interstream region around DoY 101.8 - DoY 102.5 \markme{(orange
  shading, no hatching)} is here considered as helmet-streamer
  plasma. Most of the sector-reversal regions are indeed located
  around changes in the in situ magnetic polarity; an exception is the
  period from Do{Y 9}2.1-94.1  \markme{(indicated with dark green
  shading and hatching)}, which represents slow solar wind before a
  stream interaction region without a current sheet
  crossing. \footnote{Polarity changes on scales of single or
      a few data points in the 12-minute time resolution are not
      necessarily associated with crossings of the heliospheric
      current sheet, but are most likely caused by localized kinks in
      the magnetic field \citep[]{berger2011systematic}. These
      are therefore not indications of misclassifications between helmet-streamer and sector-reversal plasma.}
In general, the sector-reversal category tends to encompass
  the slowest and densest solar wind.
 
However{, although the \citet[]{xu2014new} categorization works  reliably for the most part,} there are some {exceptions,} and a few of these are highlighted in {Fig.}~\ref{fig:timeseries}. For
example, according to the categorization, within the second high-speed
stream (~DoY 102.9 - DoY 106.2\markme{, blue shading}) two {streams of the helmet-streamer
plasma type} appear to be embedded. {Because the oxygen charge-state ratio gradually increases throughout the whole time {frame}, t}his is not very
realistic. {Fig.~\ref{fig:timeseries} also contains examples for potential misidentifications between the two slow solar wind types (helmet-streamer plasma and streamer belt plasma). W}ithin the sector-reversal stream (DoY 98 - 101.7, \markme{dark green shading, no hatching}), several short time-frames are instead categorized as
helmet-streamer plasma. Similarly, short periods of sector-reversal plasma appear to be embedded in helmet-streamer plasma (DoY
91 - DoY 91.7, \markme{orange shading with hatching})). {The latter two examples probably both contradict the concept of assigning a complete solar wind stream to the sector-reversal plasma as long as it contains a change in the magnetic field polarity. These occasional 
  misclassifications might be resolved by incorporating
  charge-state information into the solar wind classification
  method. However, this is beyond the scope of the \citet[]{xu2014new}
  scheme and this work.}


Figure~\ref{fig:cols} shows \markme{two-dimensional} histograms of the proton-proton
collisional age versus the proton speed, proton
density, proton temperature, magnetic field strength{} as a
measure for the wave activity $B_{\perp}/B${, and the
  {oxygen} charge-state ratio $\nO$}. {Each row
  corresponds to the same solar wind parameter and each column refers
  to the same solar wind type.}  The first three columns correspond to
sector-reversal plasma, helmet-streamer plasma, and coronal hole
wind. {The histogram values (indicated by the color bar) in the  first three columns are first divided by the respective bin size and
  then normalized to their corresponding maxima}. The fourth column is
the sum of the first three (normalized) histograms. This normalization
reduces the effect of different {underrepresented} solar wind
types {in the fourth column}. In each row, the proton-proton
collisional age is highest for sector-reversal plasma, has
intermediate values for helmet-streamer plasma, and is lowest for
coronal hole wind. That the collisional age is lowest for coronal hole
wind is expected. Coronal hole wind is typically
{characterized} by a low {$\nO$} charge-state ratio,
high proton speed, low proton density, and high (apparent)
{proton} temperatures. The latter three are all expressed in a
low collisional age (see Equation~\ref{eq:col}). That the collisional
age is systematically higher in sector-reversal plasma than in helmet-streamer plasma {(as is visible in
  {Fig.}~\ref{fig:cols}),}  however, is not immediately obvious
{from its definition as slow solar wind plasma that contains a
  current sheet crossing}. In the \citet[]{xu2014new} scheme, many
stream interaction boundaries are classified as sector-reversal plasma
because they (frequently but not always) contain current sheet
crossings. In compression regions in stream interaction regions, the
proton density, proton temperature, and magnetic field strength are
high (see, e.g.,
Figure~\ref{fig:timeseries}). However{,} in the compressed slow solar wind, the proton temperature is also higher than in
{uncompressed} slow solar wind. That the proton-proton
collisional age is nevertheless higher in this regime than in helmet-streamer plasma indicates that the heating in the compression regions
does not completely compensate for the increase in proton
density. As the example in {Fig.}~\ref{fig:timeseries}
illustrates, not all stream interaction regions and not only stream
interaction regions are assigned to the sector-reversal type, however. As
illustrated in {Fig.}~\ref{fig:timeseries}, the sector-reversal category also contains very slow and dense solar wind with typically
low temperatures. All three of these properties {also} lead to
a particularly high proton-proton collisional age.

For {the proton density, the proton temperature, the magnetic field strength, } and
$B_\perp/B$, the fourth column in {Fig.}~{\ref{fig:cols}} shows
three visibly separated maxima. This already provides an indication that
constant thresholds in the proton-proton collisional age alone
{(without magnetic field information)} appear to be sufficient
to distinguish between the three plasma types. This observation
motivates us to derive thresholds of constant proton-proton
collisional age as an alternative method for distinguishing between the
same solar wind types as the \cite{xu2014new} categorization scheme.

\section{{Solar wind categorized based on proton-proton collisional age }}
{In the following, we describe two alternative approaches for
  deriving threshold values in the proton-prot{o}n collisional
  age, called similarity-PAC (in this section) and distribution-PAC
  (in {Sec.}~\ref{sec:time})}.  {Both variants of the PAC
  scheme define thresholds in the proton-proton collisional age that
  aim to separate coronal hole wind, helmet-streamer plasma, and
  sector-reversal plasma.  We refer to the threshold value between
coronal hole wind and helmet-streamer plasma as {$\ac^{ch}$},
and to the threshold value between helmet-streamer plasma and sector-reversal as $\ac^{hr}$.} 

\subsection{S{imilarity-PAC} method and its relation to the \citet{xu2014new} scheme}\label{sec:ac}
As a first approach, we varied the
positions of {the} two thresholds {$\ac^{ch}$} and {$\ac^{hr}$} and
computed the similarity of the resulting proton-proton collisional
age-based classification scheme to the \citet{xu2014new} scheme. As a
similarity measure, we counted the number of data points that were
assigned to the same classes for both classification schemes and
divided this by the sample size. ICMEs were again excluded from the data
set. Figure~\ref{fig:similarity} shows this similarity measure
{for varying positions of $\ac^{ch}$ and $\ac^{hr}$}. The
highest possible value of this similarity measure is $1,$ which would
be achieved if all data points from all three considered solar wind
types were classified the same in  both classification schemes. The
$\text{plus}$  in {Fig.}~\ref{fig:similarity} indicates the highest observed similarity of $0.90$. Because the maximum is comparatively broad,
{the ranges} $\log\left(\ac^{ch}\right)\in[-1.00,-0.95]$ and
$\log\left(\ac^{hr}\right)\in[0.10,0.21]$ provide reasonable
{values} for the proposed $\ac$ method.  In this way, we can
derive thresholds for the proton-proton collisional age that result in
a categorization {that is simimlar to} the original \citet{xu2014new}
scheme. {This represents the first variant of our proposed
  solar wind classification scheme: the similarity-PAC.}

\citet{xu2014new} {provided} expressions for the decision boundaries in
terms of proton speed, proton density, proton temperature{,}
and magnetic field strength. We employed these to express the decision
boundaries in terms of the proton-proton collisional age.  The first
decision boundary (Q1, Equation 8 in \citet{xu2014new}) only separates
ejecta from the three other solar wind types. In the following{,}
we therefore only consider the remaining two decision boundaries. Expressing the
second and third decision boundaries (Q2: Equation 9 and Q3: Equation
10 in \citet{xu2014new}) in terms of the proton-proton collisional age
yields (after converting everything into {the same} units)

\begin{eqnarray*}
Q2: & \ac < & 2.542 10^{-7} \quad B^{0.4128} \Tp^{-1.209}\n^{0.3837}\\
Q3: & \ac > & 3.404 10^{-5} \quad B^{1.6899} \Tp^{0.747} \n^{-1.562} \enspace .
\end{eqnarray*}

Notably, these expressions do not directly imply a constant value of
the collisional age for each decision boundary, and they depend
explicitly on the magnetic field strength, which does not appear in
Equation~\ref{eq:col} directly. That constant thresholds in
collisional age nevertheless lead to asolar wind
classification that is very similar to the full \citet[]{xu2014new} scheme implies
{that all for this classification, required information on the
  magnetic field is already encoded in the proton plasma
  properties. This indicates an additional underlying relationship
  between the respective quantities that appears to lead to this
  effect.} For the separation between coronal hole wind and helmet-streamer plasma, this can be explained by the expected underlying
correlation between proton temperature and magnetic field strength.

\subsection{Solar cycle dependence of solar wind parameters}\label{sec:solarcycle}

\begin{figure}[H]
  \begin{center}
  \includegraphics[width=0.8\columnwidth]{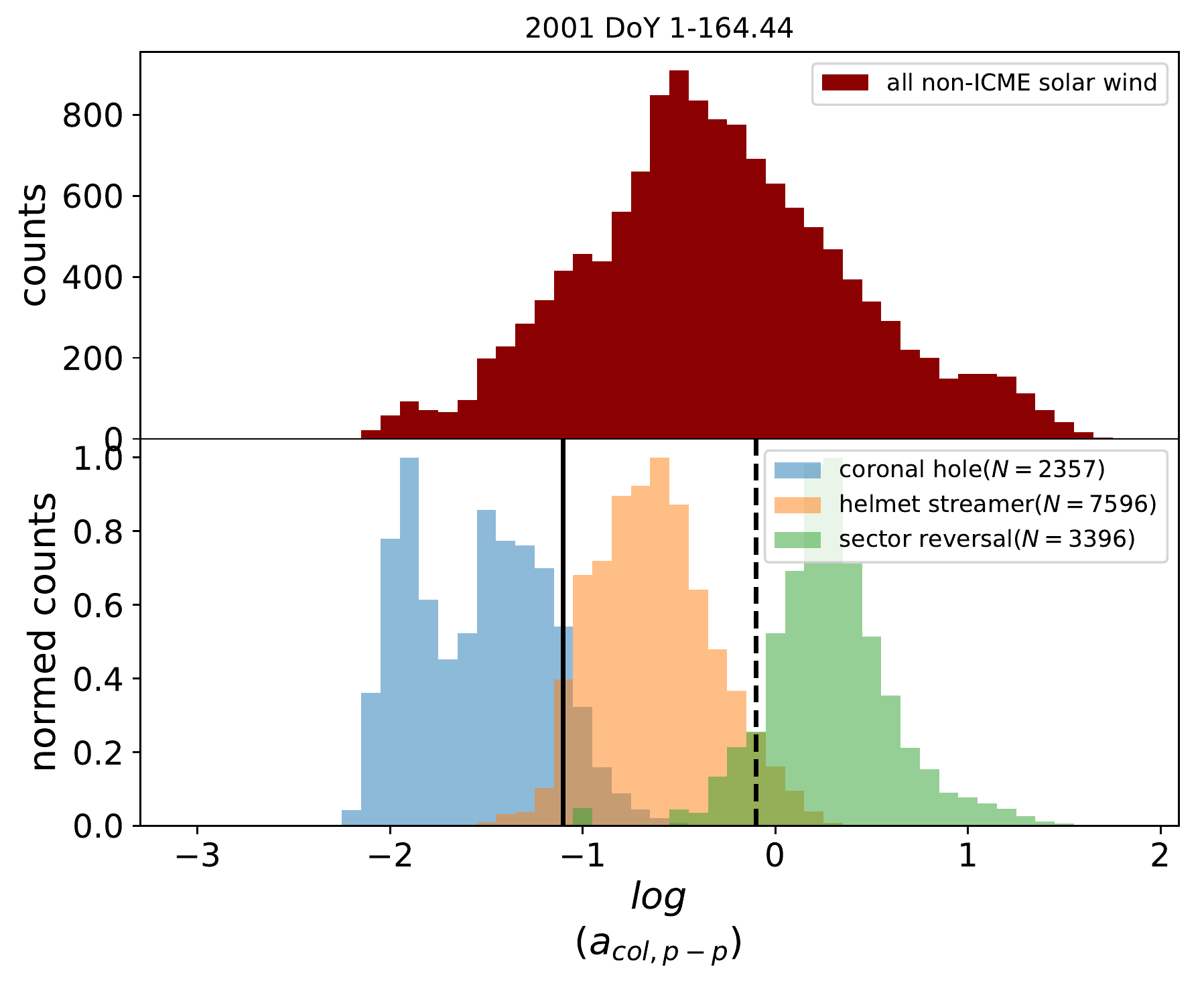}

  \includegraphics[width=0.8\columnwidth]{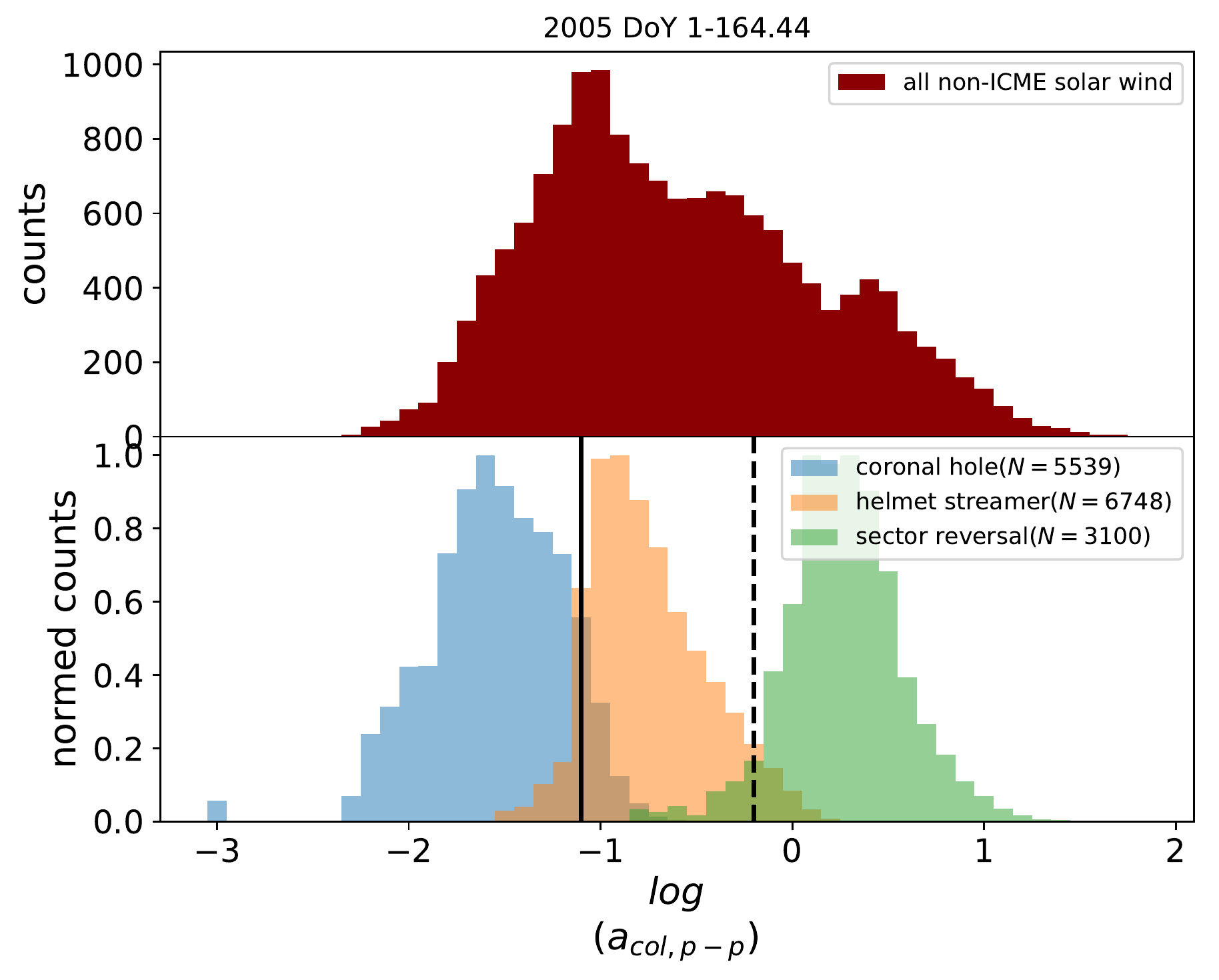}

  \includegraphics[width=0.8\columnwidth]{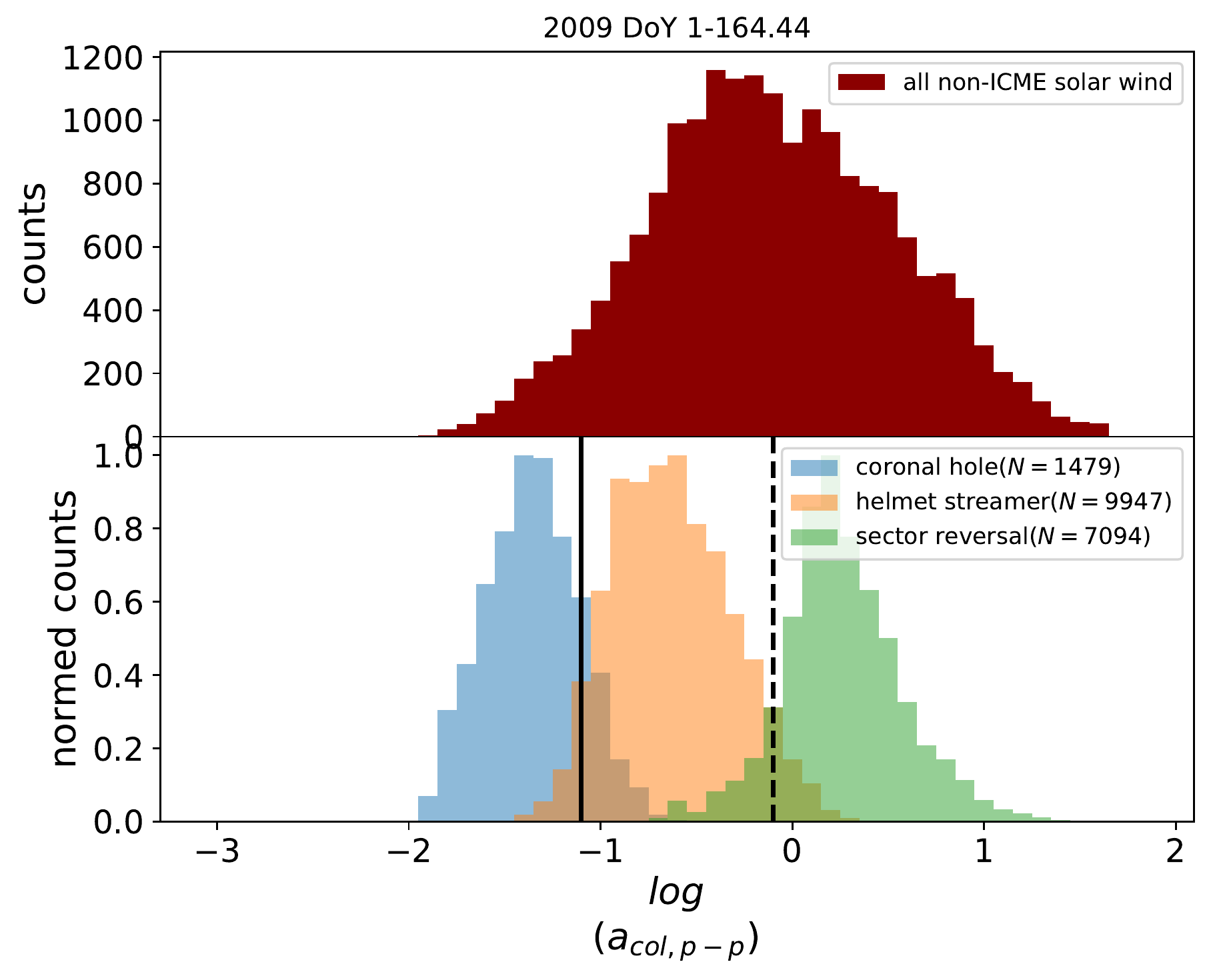}  
  \end{center}

  \caption{\label{fig:thresholds}Summed 1-dimensional histograms of the
    proton-proton collisional age for different selected time {frames}. Top: DoY 1-164.44 in 2001. Middle: DoY 1-164.44 in 2005.
    Bottom: DoY 1-164.44 in 2009. In each subplot, the upper panel
    gives the non-normalized histograms {of} all observations for the
    three solar wind types (coronal hole wind{ in dark blue}, helmet-streamer plasma {in orange,}
    and sector-reversal plasma {in dark green}, following the \citet[]{xu2014new}
    scheme), the lower panel shows the three histograms normalized
    according to solar wind type. The black solid vertical lines
    indicate the respective minimum {$\ac^{ch}$} between coronal hole
    wind and helmet-streamer plasma, and the black dashed line
    indicates the minimum {$\ac^{hr}$} between helmet-streamer and
    sector-reversal plasma. (To make all three histograms visible at
    the same time, the alpha channel is set to 0.5.) {The sample size is indicated in the legend in each subplot.}}
\end{figure}

Like the Sun itself, the solar wind properties change with the
solar activity cycle
\citep[]{manoharan2012three,lepri2013solar,zhao2010comparison,kasper2012evolution,schwadron2011coronal}. Furthermore,
\citet{zhao2014polar} showed that the coronal hole wind from equatorial
and polar coronal holes has different characteristics (independent
of the solar activity cycle). Because small equatorial coronal holes
{(at low latitudes)} are more frequent (but also more
short-lived) during solar activity maximum than during solar
activity minimum, this alone already implies different properties of a
``typical'' coronal hole wind in different phases of the solar
activity cycle. In the concept of the solar wind described by \citet{Antiochos2011}, all coronal holes are connected. Here, the differences between
polar and equatorial coronal holes can be understood as a systematic
change in solar wind properties with distance to the coronal
hole border (see \citet{peleikis2015sw14}).
\begin{figure*}
  \begin{center}
  \includegraphics[width=\textwidth]{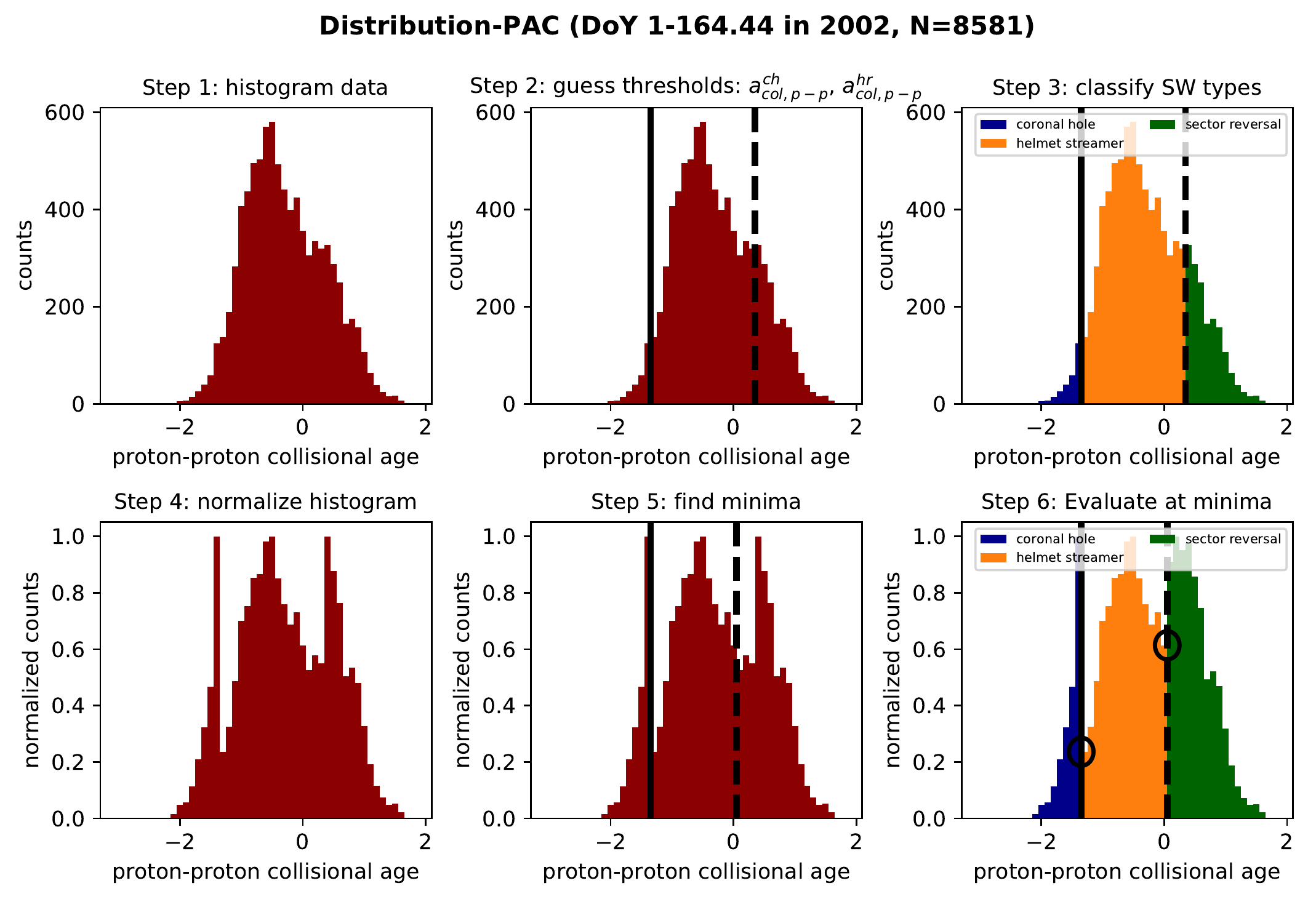}
  \end{center}

  \caption{\label{fig:method}Distribution-PAC method for one time frame (DoY 1-164.44 in 2002). \markme{Each panels shows \markme{one-dimensional} histograms of the proton-proton collisional age. In the second row, all histograms are normalized to their respective maxima. Solid and dashed vertical lines indicate classification thresholds. The circles in the last panel in the second row indicate the histogram values used for the evaluation of the candidate solution.} The sample size $N$ is given in the figure title.}
\end{figure*}

{Figure~\ref{fig:CRTS}} shows {time series} of solar wind proton
speed, proton density, proton temperature, magnetic field strength,
perpendicular variability of the magnetic field $B_\perp/B$,
Alfv{\'e}n speed, specific proton entropy, the ratio of observed
proton temperature to the expected temperature, the proton-proton
collisional age, and the monthly {sunspot} number {(from version
  2.0 of the SILSO data, Royal Observatory of Belgium, Brussels,
  \citet[]{clette2016revised})}. Except for the last panel, each panel
shows data for coronal hole wind, helmet-streamer plasma, and sector-reversal plasma, and
each data point is averaged over the entire solar wind of the respective type
from one {synodic solar rotation. We assumed an approximate length of a
  Carrington rotation of 27.24 days and used this as the bin size. The
  true start point of the Carrington rotation was ignored.} As observed
in 
\citet{manoharan2012three}, \citet{lepri2013solar}, \citet{zhao2010comparison}, \citet{kasper2012evolution}, and \citet{schwadron2011coronal},
most of these solar wind parameters change systematically following the solar activity cycle. In particular, the magnetic field strength and
the {oxygen} charge-state ratio (panels 4 and 5) exhibit lower values
during the solar activity minimum than during the solar activity
maximum. This holds for each of the three solar wind type{s} separately.
{Oxygen} charge state ratios that are associated with coronal hole wind
during the solar activity maximum would fall in the range of slow
solar wind if they were observed during the solar activity minimum. A
similar but weaker trend is also observed for proton speed, proton
temperature, and proton density. This emphasizes that solar wind
categorization based on thresholds in both charge state compositions
and proton plasma properties cannot be expected to perform equally
well in all phases of the solar activity cycle with the same fixed
threshold values. Instead, the thresholds need to be dependent on the
phase of the solar activity cycle.

For the whole time {frame} shown in {Fig.}~\ref{fig:CRTS}, the values
of the collisional age and the specific proton entropy for the three
plasma types are well separated. However, because the solar wind types
in {Fig.}~\ref{fig:CRTS} are based on the fixed \citet{xu2014new}
scheme {that does not reflect, for example, overall lower
  proton densities and magnetic field strength during the solar
  activity minimum}, the temporal variation is probably
underestimated. This also means that for the collisional age and the specific
proton entropy, the same fixed thresholds cannot (yet) be expected to
{be the optimal choice} throughout the whole solar activity cycle. The
separation between the different solar wind types shown in
{Fig.}~\ref{fig:CRTS} indicates that the specific proton entropy and
the proton-proton collisional age are best suited as single parameters
in a \markme{one-dimensional} solar wind categorization scheme. It should be
noted that a large overlap between the different solar wind types in
{Fig.}~\ref{fig:CRTS} only rules out the respective parameter for a
\markme{one-dimensional} classification scheme. The informative value of these
parameters in a higher dimensional decision space cannot be estimated
in this representation.

\subsection{\markme{Time-dependent solar wind classification based on the proton-proton collisional age}}\label{sec:time}

\begin{figure}[h]
  \includegraphics[width=\columnwidth]{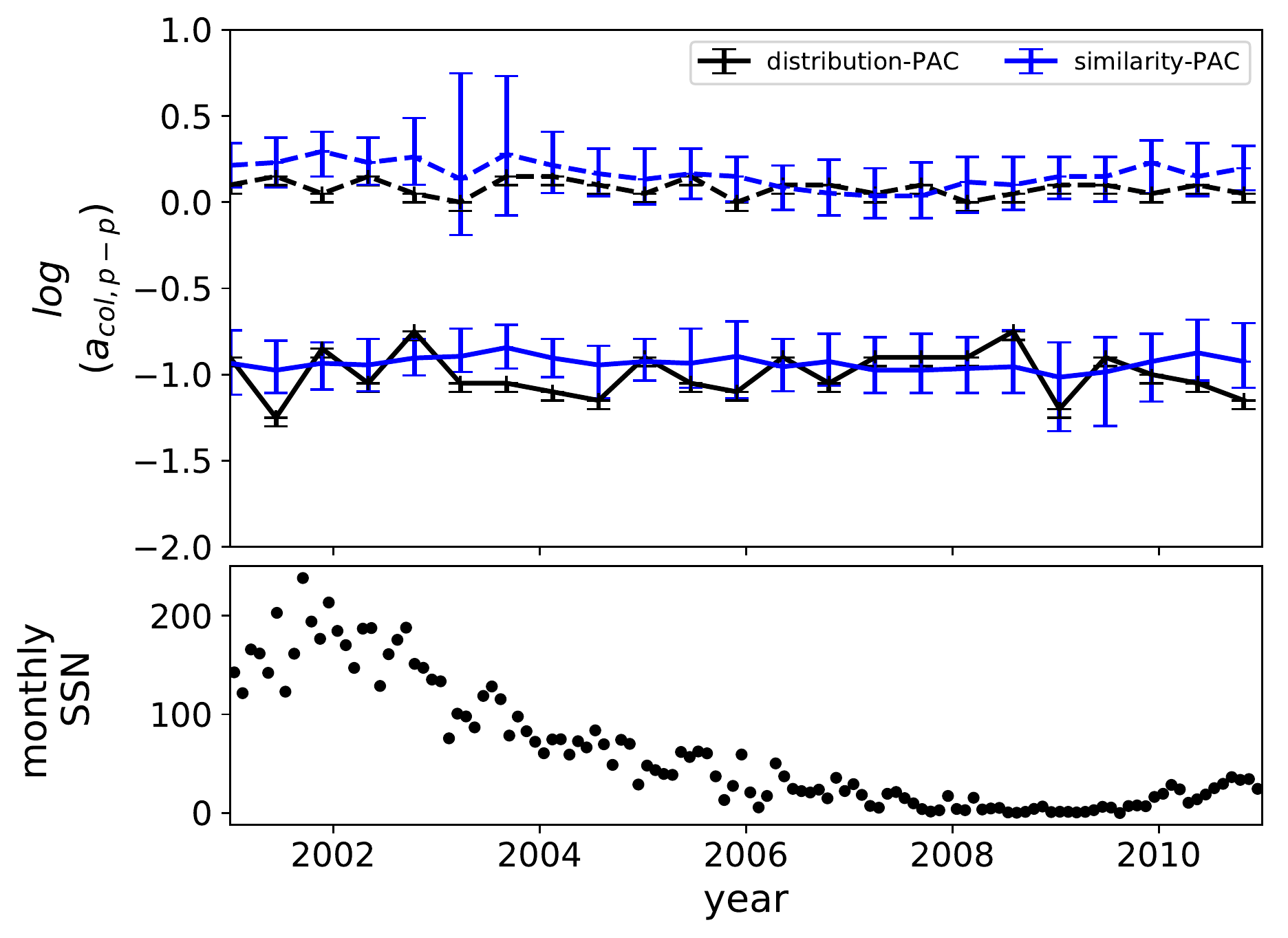}
  \caption{\label{fig:perquarter} Thresholds in proton-proton
    collisional age over time derived from {the similarity-PAC
    }(solid and dashed blue) and with {the distribution-PAC}
    (solid and dashed black). {The dashed lines indicate the
      separation threshold $\ac^{ch}$ between coronal hole wind and
      helmet-streamer plasma, and the solid lines refer to $\ac^{hr}$
      as the decision boundary between helmet-streamer and
      sector-reversal plasma.}}

\end{figure}

Based on the considerations in the previous section, we derive a
second, time-dependent criterion for threshold values in the
proton-proton collisional age. Under the assumption that all three
solar wind types are {well }represented in the data set, this
approach is based directly on the distribution of the proton-proton
collisional age {and is therefore called the distribution-PAC
  method}. {To} ensure reasonable statistics, we divided the
time series into batches of {$6\times 27.24$ days (corresponding
  to six times the approximate length of a Carrington rotation).
  Figure~\ref{fig:thresholds} shows normalized \markme{one-dimensional} histograms of the
  proton-proton collisional age for the first 163.44 days} in 2001, 2005, and 2009. To reduce the effect of
unbalanced samples for each of the three solar wind types, each solar
wind type (for illustrative purposes, the \citet{xu2014new} types are
used here again) is {binned} separately and is normalized to
its respective maximum. The lower panels {for each year} in
{Fig.}~\ref{fig:thresholds} show these normalized histograms. Then,
the minima between the two class pairs (sector-reversal plasma and
helmet-streamer plasma on the one hand, and helmet-streamer plasma and coronal hole wind on the other) are taken as the new estimate
for the threshold values. In all three subplots, a {tail} at
$\log{\ac}\sim1$ is visible. {In 2001, this population is
  frequent enough to form a third peak.} This is probably related to
ICMEs that are missing from the ICME lists or ICMEs whose start and end times are too inaccurate in the ICME lists. However, when we use the
\citet[]{xu2014new} solar wind classification as a starting point, as
in {Fig.}~\ref{fig:thresholds}, this approach requires a ground truth
for the solar wind categorization as prior information. To avoid this,
we used an iterative random search {that is illustrated in Fig.~\ref{fig:method}}: {Step 1: For each time frame a \markme{one-dimensional} histogram of the proton-proton collisional age is generated. Step 2:} An initial guess
($\log{\left(\ac^{ch}\right)}=-1$ and
{$\log{\left(\ac^{hr}\right)=0}$} modified with additive
uniform noise in $[-0.1,0.1]$) {provides candidates for the decision boundaries. Step 3: Based on these thresholds, a candidate classification is derived. Step 4: Each of these candidate solar wind types is separately normalized by its maximum. Step 5: From the sum of these normalized histograms, new thresholds are derived as the minima between the two main peaks. To increase robustness against noise and underrepresented solar wind types, the two main peaks are required to be in $\ac\in[-1.5,0.8]$. Step 6: The derived thresholds are evaluated by adding the histogram values at the minima.} These {six} steps are then repeated 10000 times. We take the result that leads to the deepest minima in
the renormalized histogram as the final solution. {This
  approach assumes that at least three solar wind types are
  always present in the data. The approach is otherwise purely data{-}driven.  This
  process is then applied to each data batch (with a length of $6\times
  27.24$ days),} and the results are shown in
{Fig.}~\ref{fig:perquarter}.

\begin{figure}[h]
  \includegraphics[width=\columnwidth]{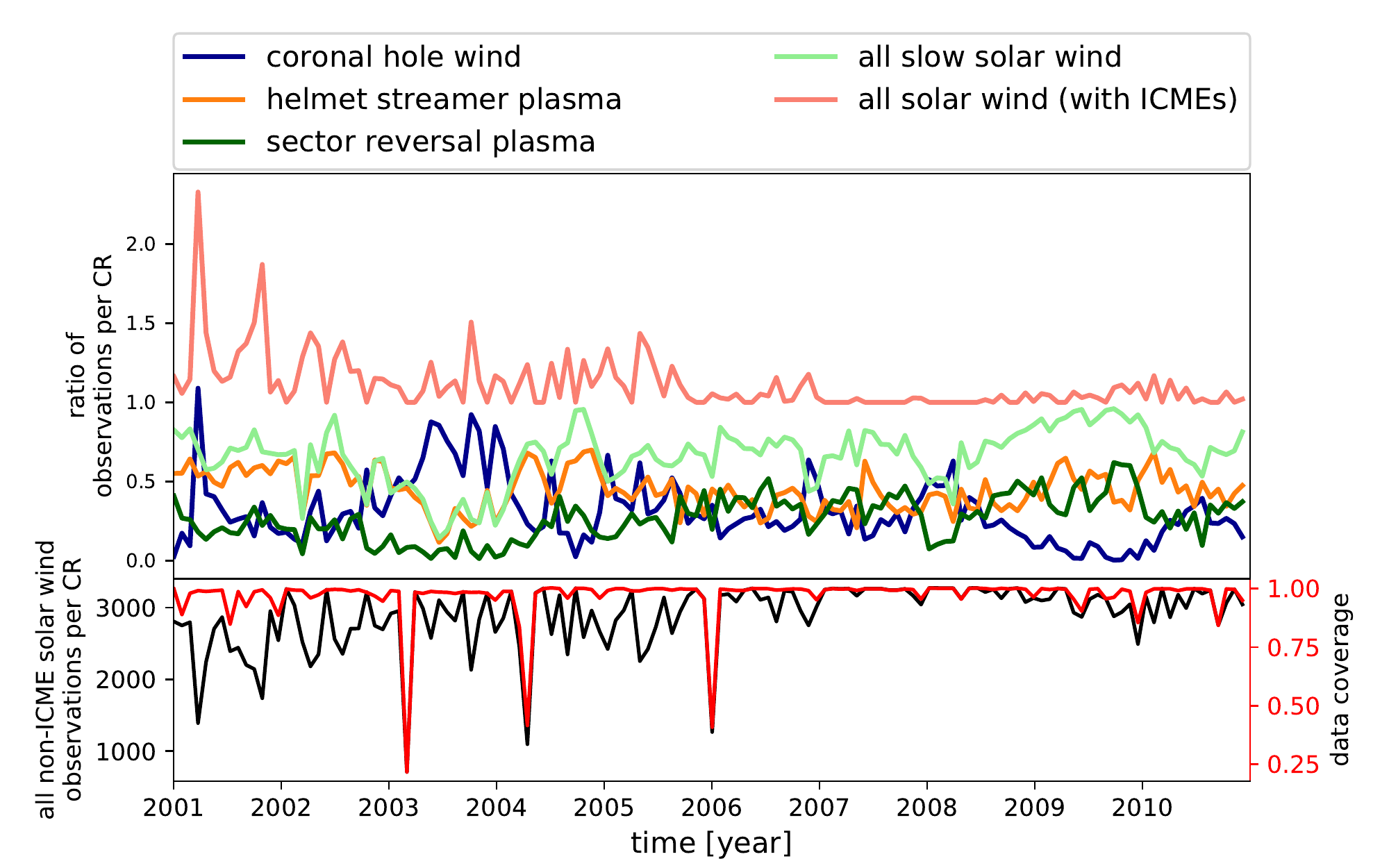}
  \caption{\label{fig:swmix} {Upper panel:} Ratio of
    observations of solar wind types (based on the \cite{xu2014new}
    scheme) relative to the entire solar wind (without ICMEs) over time
    {(in time bins of length $27.24$ days)}: the entire solar wind
    with ICMEs, coronal hole wind, helmet-streamer plasma,
    sector-reversal plasma, and the sum of the two slow solar wind
    types (entire slow solar wind). {Lower panel: Number of solar
      wind observations per time bin and data coverage.}}
\end{figure}

{In this figure, two pairs of
  threshold values are given: thresholds between coronal hole wind and
  helmet-streamer plasma (lower lines at $\log{\ac^{ch}}\sim -1$, solid
  lines), and thresholds between helmet-streamer plasma and sector-reversal plasma (upper lines at $\log{\ac^{hr}}\sim 0.1$, dashed
  lines). The blue lines {refer to the similarity-PAC method
    and} are derived based the similarity to the \citet[]{xu2014new}
  scheme, as described in {Sec.}~\ref{sec:ac}. The only difference to
  {Sec.}~\ref{sec:ac} is that the method is here also applied to the
  $6\times 27.24$ time bins. The black lines represent the results of
  the distribution-PAC approach derived from the proton-proton
  collisional age distribution as described in this section and
  illustrated in {Fig.}~\ref{fig:thresholds}.}

\begin{table*}
  \begin{center} 
    \tiny
\caption{\label{tab:thresholds}Classification thresholds derived from distribution of proton-proton collisional age in $163.44$ days time resolution.}\begin{tabularx}{\textwidth}{|l|X|X|X|X|r|}\hline
  start time & $\ac^{rh}$ & $\ac^{ch}$ & $\left(\ac^{rh}\right)_{\text{sim}}$ & $\left(\ac^{ch}\right)_{\text{sim}}$ &  min=(rel. freq.) \\\hline
2001.00 & -0.900 $\in$ [-0.950,-0.900] & 0.100 $\in$ [0.050,0.100] &  -0.934$\in$ [-1.116,-0.742]  &  0.214 $\in$ [0.032,0.406] & 0.317 \\\hline
2001.45 & -1.250 $\in$ [-1.300,-1.250] & 0.150 $\in$ [0.100,0.150] &  -0.975$\in$ [-1.106,-0.803]  &  0.230 $\in$ [0.099,0.402] & 0.383 \\\hline
2001.89 & -0.850 $\in$ [-0.900,-0.850] & 0.050 $\in$ [-0.000,0.050] &  -0.934$\in$ [-1.086,-0.813]  &  0.295 $\in$ [0.143,0.416] & 0.277 \\\hline
2002.34 & -1.050 $\in$ [-1.100,-1.050] & 0.150 $\in$ [0.100,0.150] &  -0.944$\in$ [-1.096,-0.793]  &  0.230 $\in$ [0.079,0.382] & 0.298 \\\hline
2002.79 & -0.750 $\in$ [-0.800,-0.750] & 0.050 $\in$ [-0.000,0.050] &  -0.904$\in$ [-1.005,-0.793]  &  0.263 $\in$ [0.162,0.374] & 0.179 \\\hline
2003.23 & -1.050 $\in$ [-1.100,-1.050] & -0.000 $\in$ [-0.050,-0.000] &  -0.894$\in$ [-0.985,-0.732]  &  0.133 $\in$ [0.042,0.295] & 0.079 \\\hline
2003.68 & -1.050 $\in$ [-1.100,-1.050] & 0.150 $\in$ [0.100,0.150] &  -0.843$\in$ [-0.965,-0.712]  &  0.279 $\in$ [0.158,0.410] & 0.125 \\\hline
2004.13 & -1.100 $\in$ [-1.150,-1.100] & 0.150 $\in$ [0.100,0.150] &  -0.904$\in$ [-1.015,-0.793]  &  0.214 $\in$ [0.103,0.325] & 0.330 \\\hline
2004.57 & -1.150 $\in$ [-1.200,-1.150] & 0.100 $\in$ [0.050,0.100] &  -0.944$\in$ [-1.136,-0.833]  &  0.166 $\in$ [-0.026,0.277] & 0.227 \\\hline
2005.02 & -0.900 $\in$ [-0.950,-0.900] & 0.050 $\in$ [-0.000,0.050] &  -0.924$\in$ [-1.035,-0.793]  &  0.133 $\in$ [0.022,0.265] & 0.469 \\\hline
2005.47 & -1.050 $\in$ [-1.100,-1.050] & 0.150 $\in$ [0.100,0.150] &  -0.934$\in$ [-1.076,-0.732]  &  0.166 $\in$ [0.024,0.368] & 0.668 \\\hline
2005.91 & -1.100 $\in$ [-1.150,-1.100] & -0.000 $\in$ [-0.050,-0.000] &  -0.894$\in$ [-1.136,-0.692]  &  0.149 $\in$ [-0.093,0.352] & 0.626 \\\hline
2006.36 & -0.900 $\in$ [-0.950,-0.900] & 0.100 $\in$ [0.050,0.100] &  -0.955$\in$ [-1.096,-0.793]  &  0.085 $\in$ [-0.057,0.246] & 0.620 \\\hline
2006.81 & -1.050 $\in$ [-1.100,-1.050] & 0.100 $\in$ [0.050,0.100] &  -0.924$\in$ [-1.066,-0.763]  &  0.053 $\in$ [-0.089,0.214] & 0.814 \\\hline
2007.25 & -0.900 $\in$ [-0.950,-0.900] & 0.050 $\in$ [-0.000,0.050] &  -0.975$\in$ [-1.106,-0.783]  &  0.036 $\in$ [-0.095,0.228] & 0.563 \\\hline
2007.70 & -0.900 $\in$ [-0.950,-0.900] & 0.100 $\in$ [0.050,0.100] &  -0.975$\in$ [-1.106,-0.763]  &  0.036 $\in$ [-0.095,0.248] & 0.861 \\\hline
2008.14 & -0.900 $\in$ [-0.950,-0.900] & -0.000 $\in$ [-0.050,-0.000] &  -0.965$\in$ [-1.106,-0.783]  &  0.117 $\in$ [-0.024,0.299] & 0.561 \\\hline
2008.59 & -0.750 $\in$ [-0.800,-0.750] & 0.050 $\in$ [-0.000,0.050] &  -0.955$\in$ [-1.106,-0.742]  &  0.101 $\in$ [-0.051,0.313] & 0.446 \\\hline
2009.04 & -1.200 $\in$ [-1.250,-1.200] & 0.100 $\in$ [0.050,0.100] &  -1.015$\in$ [-1.328,-0.813]  &  0.149 $\in$ [-0.164,0.352] & 0.127 \\\hline
2009.48 & -0.900 $\in$ [-0.950,-0.900] & 0.100 $\in$ [0.050,0.100] &  -0.985$\in$ [-1.298,-0.783]  &  0.149 $\in$ [-0.164,0.352] & 0.108 \\\hline
2009.93 & -1.000 $\in$ [-1.050,-1.000] & 0.050 $\in$ [-0.000,0.050] &  -0.924$\in$ [-1.157,-0.763]  &  0.230 $\in$ [-0.002,0.392] & 0.262 \\\hline
2010.38 & -1.050 $\in$ [-1.100,-1.050] & 0.100 $\in$ [0.050,0.100] &  -0.874$\in$ [-1.035,-0.682]  &  0.149 $\in$ [-0.012,0.341] & 0.625 \\\hline
2010.82 & -1.150 $\in$ [-1.200,-1.150] & 0.050 $\in$ [-0.000,0.050] &  -0.924$\in$ [-1.076,-0.702]  &  0.198 $\in$ [0.046,0.420] & 0.411 \\\hline
\end{tabularx}\end{center}\end{table*}


\begin{figure*}
  \includegraphics[width=\textwidth]{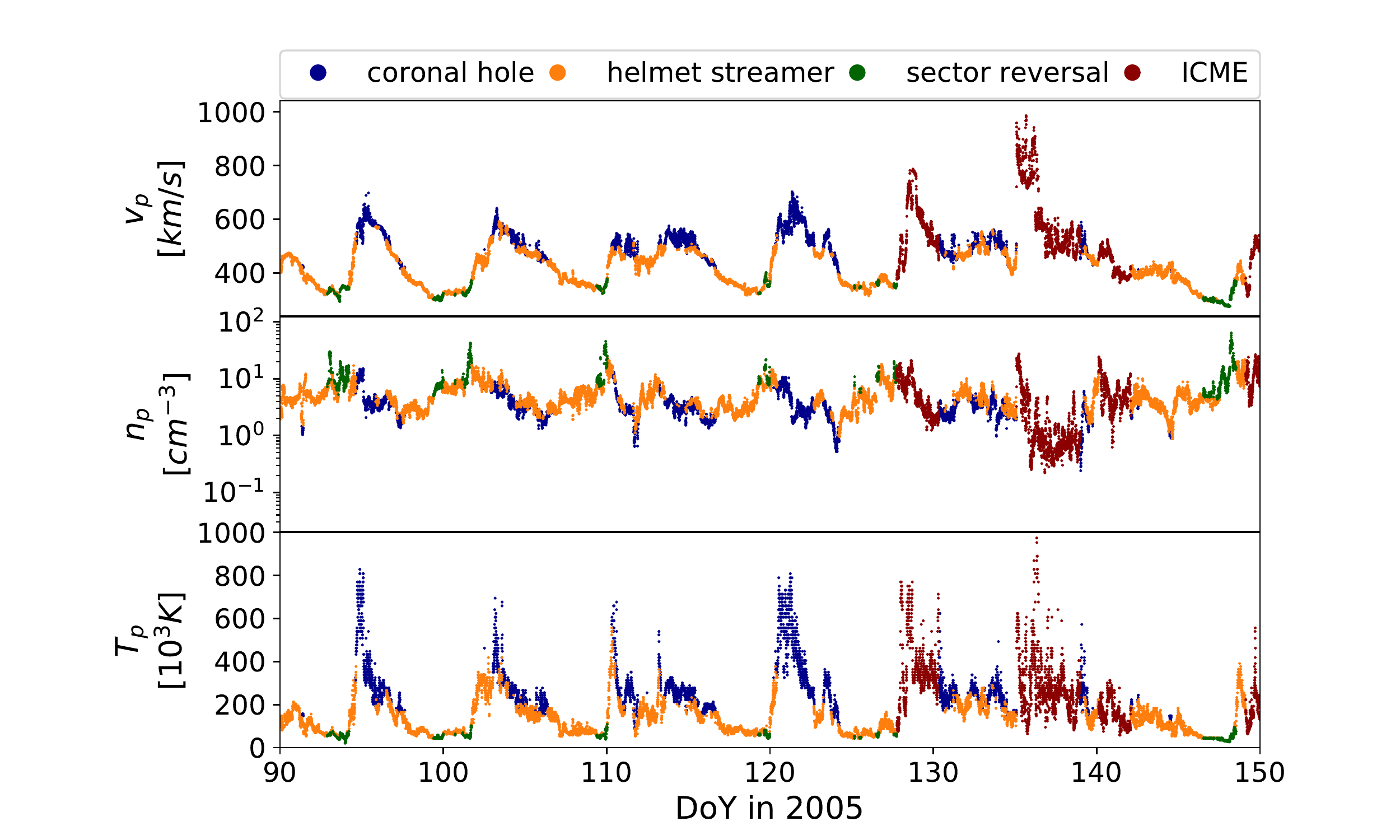}
  \caption{\label{fig:soho}Proton plasma properties from
    SOHO, CELIAS, MTOF, and PM. The colors indicate the solar wind type
    according to the distribution{-PAC}
    criterion (derived only from SOHO, CELIAS, MTOF, and PM data): dark blue
    shows the coronal hole wind, orange represents helmet-streamer plasma, and
    dark green is used for sector-reversal plasma.}
\end{figure*}

Consecutively applied to data batches, both criteria (the similarity
to the \citet[]{xu2014new} scheme and the distribution{-}based approach)
lead to {time-variable} thresholds. {However, the thresholds
  derived from both criteria (similarity-PAC and distribution-PAC)
  show comparatively small variations throughout the solar cycle,
  wherein distribution-PAC exhibits slightly larger variations. For
  the similarity-PAC this might arise because the
  \citet[]{xu2014new} scheme fixed
  separating hyperplanes throughout the solar activity cycle.}

  {The second criterion, the distribution-PAC, }directly depends on the
observed solar wind mixture. As shown in {Fig.}~\ref{fig:swmix} {(which provides an overview of the relative frequencies of the  considered solar wind types over time)}, a higher fraction of
coronal hole wind has been observed{ in 2003} than in any other year. {At} the same time, a very low fraction of sector-reversal plasma {has been observed}. In 2009{,} a very low
fraction of coronal hole wind is observed. {Both time {frames}
  violate the underlying assumption of our approach that three
  types of solar wind are always observed {with a sufficient relative
    frequency}}. {Nevertheless, distribution-PAC is able to
  provide stable classification thresholds.} {For the
  similarity-PAC and distribution-PAC,} {Fig.}~\ref{fig:perquarter}
shows no clear trend with solar activity cycle. This means that the
variability of the threshold values caused by the time-varying mix of
solar wind types is probably larger than the variability of the
proton-proton collisional age with the solar activity cycle itself.


Table~\ref{tab:thresholds} provides the threshold value for each time
{frame} from {Fig.}~\ref{fig:perquarter}. The {confidence
  intervals} (in Table~\ref{tab:thresholds} and
{Fig.}~\ref{fig:perquarter}) indicate how far the thresholds
can be shifted {in any direction} at most while changing the
{occurrence frequency of the} predicted solar wind type by at
most $5\%$ of the data. The distribution-based method relies on the
assumption that all three solar wind types are represented in any
given data set. An estimate for this is included in
Table~\ref{tab:thresholds} (last column): We take the minimum ratio of
the numbers of observations from each neighboring pair of solar wind
(coronal hole wind versus helmet-streamer plasma, and helmet-streamer
plasma versus sector-reversal plasma) in the \citet[]{xu2014new}
scheme as {an} indicator of how underrepresented the least
frequently observed solar wind type is during each time {frame}. If all
three solar wind types were equally frequent in a given data sample,
then this reliability parameter would be one. The lower this value,
the more unbalanced the data set. As expected from
{Fig.}~\ref{fig:swmix}, in 2003 and 2009, the data set
suffers most from underrepresentation (sector-reversal plasma is
underrepresented in 2003, and coronal hole wind is
underrepresented in 2009).

\subsection{{Proton plasma solar wind classification without a magnetometer: the distribution-PAC method applied to SOHO}}
As long as the assumption is valid that the observed solar wind covers
all three considered solar wind types, \markme{the proposed distribution-PAC approach} can be applied to other
instruments (and other positions in the heliosphere). As a test case,
we applied this method to the plasma data obtained from the proton
monitor (PM). The PM is an auxiliary instrument to the mass
time-of-flight (MTOF) sensor that is part of the charge, element, and
isotope analysis system (CELIAS, \citet[]{Hovestadt-etal-1995})
on board the Solar and Heliospheric Observatory (SOHO). {We
  chose SOHO as a test case because both SOHO and ACE are located close
  to L1 during the time {frame} shown in {Fig.}~\ref{fig:timeseries}
  and {Fig.}~\ref{fig:soho}. Therefore we can expect that both  spacecraft observer similar solar wind streams. Because SOHO is not
  equipped with a magnetometer, the full \citet[]{xu2014new} scheme
  cannot be applied directly.} An example {of} the resulting solar
wind classification is shown in {Fig.}~\ref{fig:soho}. As the for full
\citet[]{xu2014new} scheme based on ACE data for the same time {frame}
(see {Fig.}~\ref{fig:timeseries}), this simple method also leads to
some {misclassifications}. For the most part, however, the solar wind
classification behaves as expected: Sector-reversal plasma is mainly
concentrated before and during stream interaction regions, and hot,
fast, and {thin} solar wind streams are categorized as coronal hole
wind.

\section{Discussion and conclusion}
We find that the proton-proton collisional age alone is similarly well
suited to identify the solar wind type {as} the full
\citet{xu2014new} scheme. This is not {a direct consequence} of the decision boundaries from \citet{xu2014new}, but implies an
underlying relationship between the magnetic field strength and the proton
plasma properties in the solar wind.  This relationship is an inherent
consequence of the different plasma properties in coronal hole wind
and slow solar wind. Typical coronal hole wind
properties result in a low proton-proton collisional age, whereas the
typical properties in the slow solar wind are expressed in a higher
proton-proton collisional age. Although not shown here, ICMEs are also
typically associated with high {proton-proton} collisional ages (because the proton temperature tends to be particularly low and the
proton density high). However, from our point of view, ICMEs cannot be
separated from sector-reversal plasma with the {proton-proton}
collisional age alone. In general, context information is required to
correctly and uniquely identify ICMEs. Outside of ICMEs, sector-reversal plasma (which includes compressed slow solar wind around
current sheet crossings) shows the highest proton-proton collisional
age. These regions are also characterized by a higher magnetic field
strength. This is not directly encoded in the proton-proton collisional age. \markme{That the proposed simplified classification scheme based on the proton-proton collisional age}
{(PAC)}  nevertheless leads to
a classification that is very similar to the full \citet[]{xu2014new} scheme
(which has decision boundaries that depend on the magnetic field
strength) indicates that all required information about the magnetic
field strength is already encoded in the proton plasma properties.

\markme{The clear separation visible in the proton-proton collisional age
between coronal hole wind and slow solar wind} is probably sharpened by
the respective presence and absence of wave activity. Wave activity
together with a high Alfv{\'e}n velocity in coronal hole wind leads to
(apparent) heating of the solar wind and thus to a further decrease
in proton-proton collisional age. That the proton-proton
collisional age also separates helmet-streamer and sector-reversal
plasma is probably a combination of two effects: the sector-reversal
plasma tends to include the densest and slowest slow solar wind
streams and in addition contains most of the slow solar wind
compression regions. Both types of plasma are characterized by a
particularly high collisional age. The collisional age is interpreted as
a measure of the history of a particular solar wind plasma package. Computing {the proton-proton collisional age} based on the in situ proton plasma
properties implies the assumption that these conditions affect the
solar wind during its complete travel time. However, this assumption
is violated in stream interaction regions that become increasingly more
frequent and extended farther away from the Sun. In this case,
the proton-proton collisional {age} as considered here therefore overestimates the
effect of collisions. For the sake of categorizing the solar wind based
on the proton-proton collisional age, this turns out to be beneficial
because it increases the differences between helmet-streamer and
sector-reversal plasma. {The collisional thermalization of the
  solar wind as described in \citet[]{maruca2013collisional} presents
  an alternative that avoids this ambiguity. Instead of assuming fixed
  values for the solar wind proton speed, density, and temperature,
  \citet[]{maruca2013collisional} allowed for a radial evolution of
  these quantities. However, this requires a model for the radial
  evolution.}

{We proposed two variants of our categorization of the solar wind based on proton-proton collisional
  age: similarity-PAC, which is based
  on the observed similarity to the \citet[]{xu2014new} scheme, and
  distribution-PAC, which is a purely data-driven categorization of the
  solar wind that only relies on the assumption that sufficiently
  frequent observations from the three solar wind types are contained
  in the data.}

Because the Sun induces systematic changes in the solar wind properties
during its solar activity cycle, any solar wind categorization scheme
depending on solar wind composition or proton plasma properties should
reflect this by featuring thresholds that depend on the solar cycle. Nevertheless, for our {PAC schemes}, the derived
thresholds are more sensitive to the time-varying mixture of solar
wind types than to {systematic changes in solar wind
  properties of each solar wind type with} the solar activity cycle.

An advantage of the proposed collisional age approach is that the
categorization scheme does not require magnetic field measurements,
but only proton speed, proton density, and proton temperature. Unlike
the full scheme of Yu \& Borovsky (2015), the PAC method can be directly applied to observations from the proton
monitor on SOHO.  A drawback
of this approach (and all solar wind categorization schemes that are based on
proton plasma properties in general) is that the separating
hyperplanes depend on radius and therefore can only be applied at 1 AU in their current form.

\begin{acknowledgements}
      Part of this work was supported by the Deut\-sche
      For\-schungs\-ge\-mein\-schaft (DFG)\/ project number
      Wi-2139/11-1\enspace.  We further thank the science teams of
      ACE/SWEPAM, ACE/MAG as well as ACE/SWICS for providing the
      respective level 2 and level 1 data products. We gratefully acknowledge the diligent and detailed work of the referee.  
\end{acknowledgements}
%
   \bibliographystyle{aa} 
   \bibliography{37378_final} 

\begin{thebibliography}{35}
\expandafter\ifx\csname natexlab\endcsname\relax\def\natexlab#1{#1}\fi

\bibitem[{Antiochos {et~al.}(2011)Antiochos, Miki{\'c}, Titov, Lionello, \&
  Linker}]{Antiochos2011}
Antiochos, S., Miki{\'c}, Z., Titov, V., Lionello, R., \& Linker, J. 2011, The
  Astrophysical Journal, 731, 112

\bibitem[{Bale {et~al.}(2019)Bale, Badman, Bonnell, Bowen, Burgess, Case,
  Cattell, Chandran, Chaston, Chen, {et~al.}}]{bale2019highly}
Bale, S., Badman, S., Bonnell, J., {et~al.} 2019, Nature, 1

\bibitem[{Berger(2008)}]{berger2008velocity}
Berger, L. 2008, PhD thesis, Kiel, Christian-Albrechts-Universit{\"a}t, Diss.,
  2008

\bibitem[{Berger {et~al.}(2011)Berger, Wimmer-Schweingruber, \&
  Gloeckler}]{berger2011systematic}
Berger, L., Wimmer-Schweingruber, R., \& Gloeckler, G. 2011, Physical Review
  Letters, 106, 151103

\bibitem[{Camporeale {et~al.}(2017)Camporeale, Car{\`e}, \&
  Borovsky}]{camporeale2017classification}
Camporeale, E., Car{\`e}, A., \& Borovsky, J.~E. 2017, Journal of Geophysical
  Research: Space Physics, 122, 10

\bibitem[{Cane \& Richardson(2003)}]{cane2003interplanetary}
Cane, H. \& Richardson, I. 2003, Journal of Geophysical Research: Space
  Physics, 108

\bibitem[{Clette {et~al.}(2016)Clette, Lef{\`e}vre, Cagnotti, Cortesi, \&
  Bulling}]{clette2016revised}
Clette, F., Lef{\`e}vre, L., Cagnotti, M., Cortesi, S., \& Bulling, A. 2016,
  Solar Physics, 291, 2733

\bibitem[{D{$^\prime$}Amicis \& Bruno(2015)}]{d2015origin}
D{$^\prime$}Amicis, R. \& Bruno, R. 2015, The Astrophysical Journal, 805, 84

\bibitem[{Gloeckler {et~al.}(1998)Gloeckler, Cain, Ipavich, Tums, Bedini, Fisk,
  Zurbuchen, Bochsler, Fischer, Wimmer-Schweingruber,
  {et~al.}}]{gloeckler-etal-1998}
Gloeckler, G., Cain, J., Ipavich, F., {et~al.} 1998, in The Advanced
  Composition Explorer Mission (Springer), 497--539

\bibitem[{Heidrich-Meisner \& Wimmer-Schweingruber(2018)}]{heidrich2018solar}
Heidrich-Meisner, V. \& Wimmer-Schweingruber, R.~F. 2018, in Machine learning
  techniques for space weather (Elsevier), 397--424

\bibitem[{Hovestadt {et~al.}(1995)Hovestadt, Hilchenbach, B{\"u}rgi, Klecker,
  Laeverenz, Scholer, Gr{\"u}nwaldt, Axford, Livi, Marsch,
  {et~al.}}]{Hovestadt-etal-1995}
Hovestadt, D., Hilchenbach, M., B{\"u}rgi, A., {et~al.} 1995, Solar Physics,
  162, 441

\bibitem[{Jian {et~al.}(2011)Jian, Russell, \& Luhmann}]{jian2011comparing}
Jian, L., Russell, C., \& Luhmann, J. 2011, Solar Physics, 274, 321

\bibitem[{Jian {et~al.}(2006)Jian, Russell, Luhmann, \&
  Skoug}]{jian2006properties}
Jian, L., Russell, C., Luhmann, J., \& Skoug, R. 2006, Solar Physics, 239, 393

\bibitem[{Kasper {et~al.}(2019)Kasper, Bale, Belcher, Berthomier, Case,
  Chandran, Curtis, Gallagher, Gary, Golub, {et~al.}}]{kasper2019alfvenic}
Kasper, J., Bale, S., Belcher, J., {et~al.} 2019, Nature, 1

\bibitem[{Kasper {et~al.}(2008)Kasper, Lazarus, \& Gary}]{kasper2008hot}
Kasper, J., Lazarus, A., \& Gary, S. 2008, Physical review letters, 101, 261103

\bibitem[{Kasper {et~al.}(2012)Kasper, Stevens, Korreck, Maruca, Kiefer,
  Schwadron, \& Lepri}]{kasper2012evolution}
Kasper, J., Stevens, M., Korreck, K., {et~al.} 2012, The Astrophysical Journal,
  745, 162

\bibitem[{Lepri {et~al.}(2013)Lepri, Landi, \& Zurbuchen}]{lepri2013solar}
Lepri, S., Landi, E., \& Zurbuchen, T. 2013, The Astrophysical Journal, 768, 94

\bibitem[{Lin {et~al.}(1995)Lin, Anderson, Ashford, Carlson, Curtis, Ergun,
  Larson, McFadden, McCarthy, Parks, {et~al.}}]{lin1995three}
Lin, R., Anderson, K., Ashford, S., {et~al.} 1995, Space Science Reviews, 71,
  125

\bibitem[{Manoharan(2012)}]{manoharan2012three}
Manoharan, P. 2012, The Astrophysical Journal, 751, 128

\bibitem[{Marsch {et~al.}(1982)Marsch, M{\"u}hlh{\"a}user, Schwenn, Rosenbauer,
  Pilipp, \& Neubauer}]{marsch1982solar}
Marsch, E., M{\"u}hlh{\"a}user, K.-H., Schwenn, R., {et~al.} 1982, Journal of
  Geophysical Research: Space Physics, 87, 52

\bibitem[{Maruca {et~al.}(2013)Maruca, Bale, Sorriso-Valvo, Kasper, \&
  Stevens}]{maruca2013collisional}
Maruca, B.~A., Bale, S.~D., Sorriso-Valvo, L., Kasper, J.~C., \& Stevens, M.~L.
  2013, Physical review letters, 111, 241101

\bibitem[{McComas {et~al.}(1998)McComas, Bame, Barker, Feldman, Phillips,
  Riley, \& Griffee}]{mccomas1998solar}
McComas, D., Bame, S., Barker, P., {et~al.} 1998, in The Advanced Composition
  Explorer Mission (Springer), 563--612

\bibitem[{McComas {et~al.}(2000)McComas, Barraclough, Funsten, Gosling,
  Santiago-Mu{\~n}oz, Skoug, Goldstein, Neugebauer, Riley, \&
  Balogh}]{mccomas2000solar}
McComas, D., Barraclough, B., Funsten, H., {et~al.} 2000, Journal of
  Geophysical Research: Space Physics, 105, 10419

\bibitem[{{Peleikis} {et~al.}(2016){Peleikis}, {Kruse}, {Berger}, {Drews}, \&
  {Wimmer-Schweingruber}}]{peleikis2015sw14}
{Peleikis}, T., {Kruse}, M., {Berger}, L., {Drews}, C., \&
  {Wimmer-Schweingruber}, R.~F. 2016, in American Institute of Physics
  Conference Series, Vol. 1720, American Institute of Physics Conference
  Series, 020003

\bibitem[{Richardson \& Cane(2010)}]{richardson2010near}
Richardson, I. \& Cane, H. 2010, Solar Physics, 264, 189

\bibitem[{Sanchez-Diaz {et~al.}(2016)Sanchez-Diaz, Rouillard, Lavraud, Segura,
  Tao, Pinto, Sheeley, \& Plotnikov}]{sanchez2016very}
Sanchez-Diaz, E., Rouillard, A.~P., Lavraud, B., {et~al.} 2016, Journal of
  Geophysical Research: Space Physics, 121, 2830

\bibitem[{Schwadron {et~al.}(2011)Schwadron, Smith, Spence, Kasper, Korreck,
  Stevens, Maruca, Kiefer, Lepri, \& McComas}]{schwadron2011coronal}
Schwadron, N., Smith, C., Spence, H.~E., {et~al.} 2011, The Astrophysical
  Journal, 739, 9

\bibitem[{Smith {et~al.}(1998)Smith, L’Heureux, Ness, Acu{\~n}a, Burlaga, \&
  Scheifele}]{smith1998ace}
Smith, C.~W., L’Heureux, J., Ness, N.~F., {et~al.} 1998, in The Advanced
  Composition Explorer Mission (Springer), 613--632

\bibitem[{Stakhiv {et~al.}(2015)Stakhiv, Landi, Lepri, Oran, \&
  Zurbuchen}]{stakhiv2015origin}
Stakhiv, M., Landi, E., Lepri, S.~T., Oran, R., \& Zurbuchen, T.~H. 2015, The
  Astrophysical Journal, 801, 100

\bibitem[{Tracy {et~al.}(2016)Tracy, Kasper, Raines, Shearer, Gilbert, \&
  Zurbuchen}]{tracy2016constraining}
Tracy, P.~J., Kasper, J.~C., Raines, J.~M., {et~al.} 2016, Physical review
  letters, 116, 255101

\bibitem[{{von Steiger} {et~al.}(2000){von Steiger}, Schwadron, Fisk, Geiss,
  Gloeckler, Hefti, Wilken, Wimmer-Schweingruber, \&
  Zurbuchen}]{vonSteiger2000}
{von Steiger}, R., Schwadron, N., Fisk, L., {et~al.} 2000, {J}ournal of
  {G}eophysical {R}esearch, 105, 27

\bibitem[{Wilson~III {et~al.}(2018)Wilson~III, Stevens, Kasper, Klein, Maruca,
  Bale, Bowen, Pulupa, \& Salem}]{wilson2018statistical}
Wilson~III, L.~B., Stevens, M.~L., Kasper, J.~C., {et~al.} 2018, The
  Astrophysical Journal Supplement Series, 236, 41

\bibitem[{Xu \& Borovsky(2015)}]{xu2014new}
Xu, F. \& Borovsky, J.~E. 2015, Journal of Geophysical Research: Space Physics,
  120, 70

\bibitem[{Zhao \& Fisk(2010)}]{zhao2010comparison}
Zhao, L. \& Fisk, L. 2010, in SOHO-23: Understanding a Peculiar Solar Minimum,
  Vol. 428, 229

\bibitem[{Zhao \& Landi(2014)}]{zhao2014polar}
Zhao, L. \& Landi, E. 2014, The Astrophysical Journal, 781, 110

\end{thebibliography}
%

\end{document}